\documentclass[12pt,preprint]{aastex}

\slugcomment{To appear in \apj}

\shorttitle{The most massive core collapse supernova progenitors}
\shortauthors{Waldman}

\begin{document}


\title{The most massive core collapse supernova progenitors}


\author{R. Waldman}
\affil{Racah Institute of Physics, The Hebrew University, Jerusalem
91904, Israel}
\email{waldman@cc.huji.ac.il}


\begin{abstract}
The discovery of the extremely luminous supernova SN~2006gy,
possibly interpreted as a pair instability supernova, renewed the
interest in very massive stars. We explore the evolution of these
objects, which end their life as pair instability supernovae or as
core collapse supernovae with relatively massive iron cores, up to
about $3\,M_\odot$.
\end{abstract}

\keywords{stars: evolution, (stars:) supernovae: general}

\section{Introduction}
The interest in the evolution of very massive stars (VMS), with
masses $\gtrsim 100\,M_\odot$, has recently been revived by the
discovery of SN~2006gy - the most luminous supernova ever recorded
\citep{Ofek2007ApJ...659L..13O, Smith2007ApJ...666.1116S}. This
object, having a luminosity of $\sim 10$ times that of a typical
core-collapse SN (CCSN), is probably the first evidence of a pair
instability SN (PISN) \cite{Woosley2007Natur.450..390W}. PISN are
massive stellar objects, whose evolutionary path brings their center
into a region in thermodynamical phase space $(\rho \lesssim 10^6, T
\gtrsim 10^9)$, where thermal energy is converted into the
production of electron-positron pairs, thus resulting in loss of
pressure and hydrodynamic instability. This type of supernova was
first suggested 40 years ago by
\cite{Rakavy&Shaviv1967ApJ...148..803R,
Barkat1967PhysRevLett.18.379}, and since then several works were
carried out \cite[e.g.][]{Fraley1968Ap&SS...2...96F,
Ober1983A&A...119...61O, Ober1983A&A...119...54E,
Bond1984ApJ...280..825B, Heger2002ApJ...567..532H,
Hirschi2004A&A...425..649H, Eldridge&Tout2004MNRAS.353...87E,
Nomoto2005ASPC..332..374N}, however the overall interest in this
topic has been relatively small, mainly due to lack of observational
data.

It was originally believed that stars massive enough to produce PISN
could only be found among population III stars with close to zero
metallicity ($Z \lesssim 10^{-4}$), and hence only at very high
redshift ($z \gtrsim 15$). More recently
\cite{Scannapieco2005ApJ...633.1031S} discussed the detectability of
PISN at redshift of $z \leq 6$, arguing that metal enrichment is a
local process, therefore metal-free star-forming pockets may be
found at such low redshifts. \cite{Langer2007A&A...475L..19L}
introduced the effect of rotation into studying this question
concluding that PISN could be produced by slow rotators of
metallicity $Z \lesssim Z_\odot/3$ at a rate of one in every 1000 SN
in the local universe. Furthermore, \cite{Smith2007ApJ...666.1116S}
point out, that mass loss rates in the local universe might be much
lower than previously thought, so that massive stars might be left
with enough mass to become PISN. This conclusion is also supported
by \cite{Yungelson2008A&A...477..223Y} who extensively discuss the
mass loss rates and fates of VMS. It is interesting to note, that
SN~2006gy took place in the nearby Universe. Following the discovery
of SN~2006gy, \cite{Umeda&Nomoto2008ApJ...673.1014U} addressed the
question of how much $^{56}Ni$ can be produced in massive CCSN,
while \cite{Heger&Woosley2008arXiv0803.3161H} computed the detailed
nucleosynthesis in these SNe.

The interest in VMS is further motivated by the discovery of
Ultraluminous X-ray Sources (ULX), which can be interpreted as
mass-accreting intermediate mass black holes (IMBH) with mass $\sim
(10^2 - 10^5)\,M_\odot$. One of the possible scenarios for IMBH
formation is by VMS formed by stellar mergers in compact globular
clusters \cite[see e.g.][and references
therein]{Yungelson2008A&A...477..223Y}. In this context,
\cite{Nakazato2006ApJ...645..519N, Nakazato2007ApJ...666.1140N}
studied the collapse of massive iron cores with $M \gtrsim
3\,M_\odot$. In their first paper they treat the fate of stars of
mass $\geq 300\,M_\odot$ which reach the photodisintegration
temperature ($\approx 6 \times 10^9 K$) after undergoing pair
instability. The entropy per baryon of these models at
photodisintegration is $s>16k_B$ compared with the classical
core-collapse SN with $s \sim 1k_B$. In the second paper they aim to
bridge this entropy gap, corresponding to core masses of $(3 -
30)\,M_\odot$ but claim that there is a lack of systematic
progenitor models for this range, hence they use synthetic initial
models for their calculations.

In this work we focus mostly on the mass range $M \lesssim
80\,M_\odot$ (He core mass $M_{He} \lesssim 36\,M_\odot$)
immediately below the range which enters the pair instability
region, and present a systematic picture of the resulting CCSN
progenitors.

\section{Method}

Since the mass loss rates of stars in this range are highly
uncertain, \cite[see e.g. discussion
by][]{Yungelson2008A&A...477..223Y}, we avoid dealing with this
question by following the example of
\cite{Heger2002ApJ...567..532H}, and modeling the evolution of
helium cores. Our helium core initial models are homogeneous
polytropes composed entirely of helium and metals, with metallicity
$Z \approx 0.015$, in the mass range $(8-160)\,M_\odot$ . The models
were then evolved to the helium zero age main sequence. In the
following we will refer to these models as ``He\textit{N}'' where
\textit{N} is the mass of the model. For comparison we evolved also
a few models of regular hydrogen stars, beginning from the zero-age
main sequence (ZAMS). We will refer to these models as
``M\textit{N}'' where \textit{N} is the mass of the model. All our
models have no mass-loss. We argue that as long as the mass loss
rate is not so high that it will cut into the He-core, the evolution
after the main-sequence phase will be virtually independent of the
fate of the hydrogen-rich envelope. We followed the evolution of
each model until the star is either completely disrupted (for the
PISN case) or Fe begins to photo-disintegrate (for the CCSN case).

We followed the evolution using the Lagrangian one dimensional Tycho
evolutionary code version 6.92 (with some modifications), publicly
available on the web \cite[the code is described
in][]{Young_Arnett2005ApJ}. Convection is treated using the well
known mixing length theory (MLT) with the Ledoux criterion. In the
MLT formulation of Tycho, the value of the mixing length parameter
fit to the Sun is $\alpha_{MLT} \approx 2.1$
\citep{Young_Arnett2005ApJ}, so we used a value of
$\alpha_{MLT} = 2$ in our calculations. The nuclear reaction rates
used by TYCHO are taken from the NON-SMOKER database as described in
\cite{Rauscher_Thielemann2000ADNDT..75....1R}.

The evolution is generally followed using the code's hydrostatic
mode. The pulsational pair instability models are treated as
follows. When hydrodynamic instability is encountered, the code is
switched to the hydrodynamic mode, and mass ejection is accounted
for by removing outer zones having supersonic velocity in excess of
the escape velocity. After mass ejection has died out, and the
stellar core is already in contraction, the code is switched back to
the hydrostatic mode to follow the interpulse period.

\section{Results}

\begin{figure}
\begin{minipage}[t]{0.5\linewidth}
\centering
\includegraphics[width=1.\linewidth]{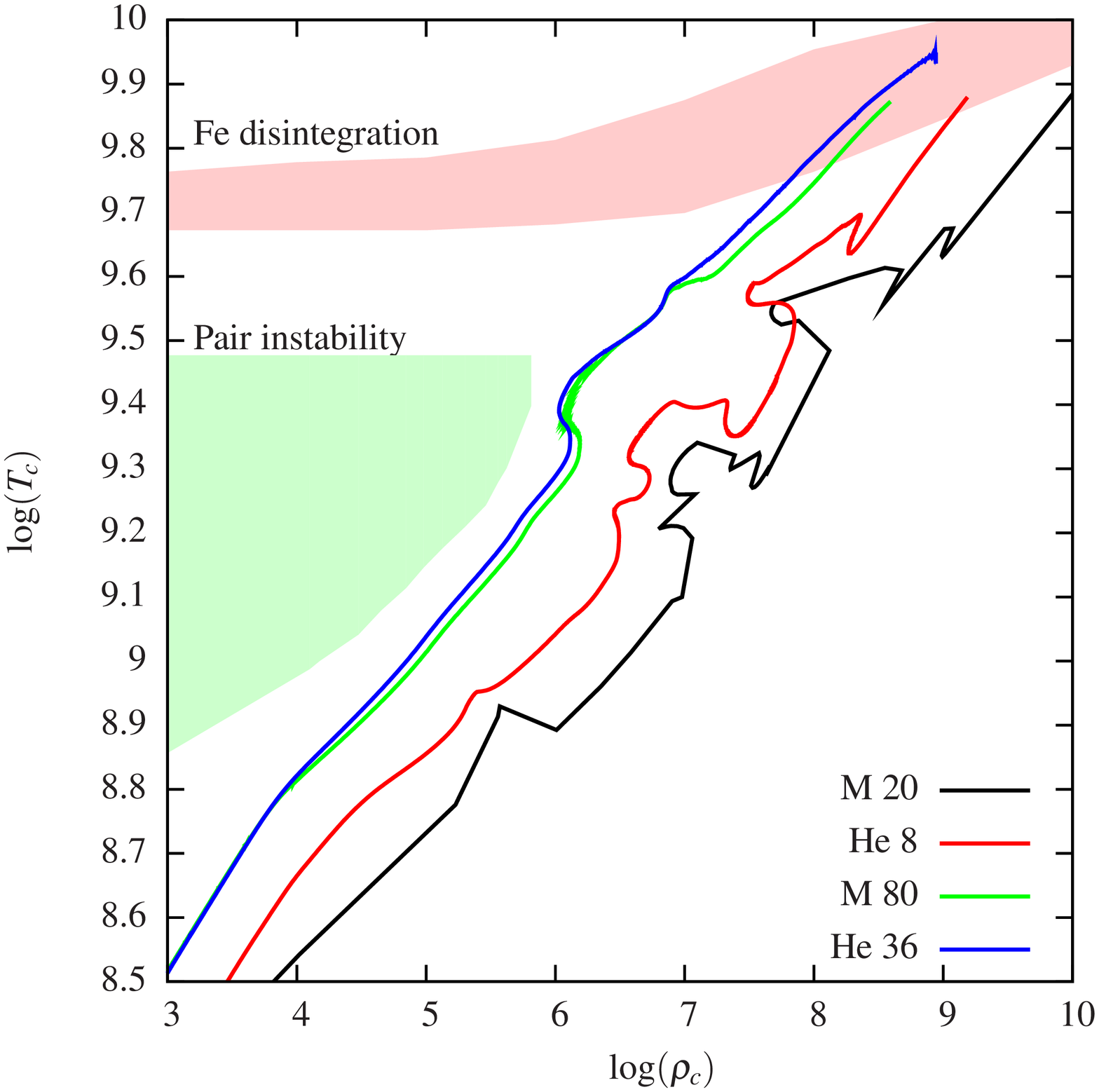}
\end{minipage}
\begin{minipage}[t]{0.5\linewidth}
\centering
\includegraphics[width=1.\linewidth]{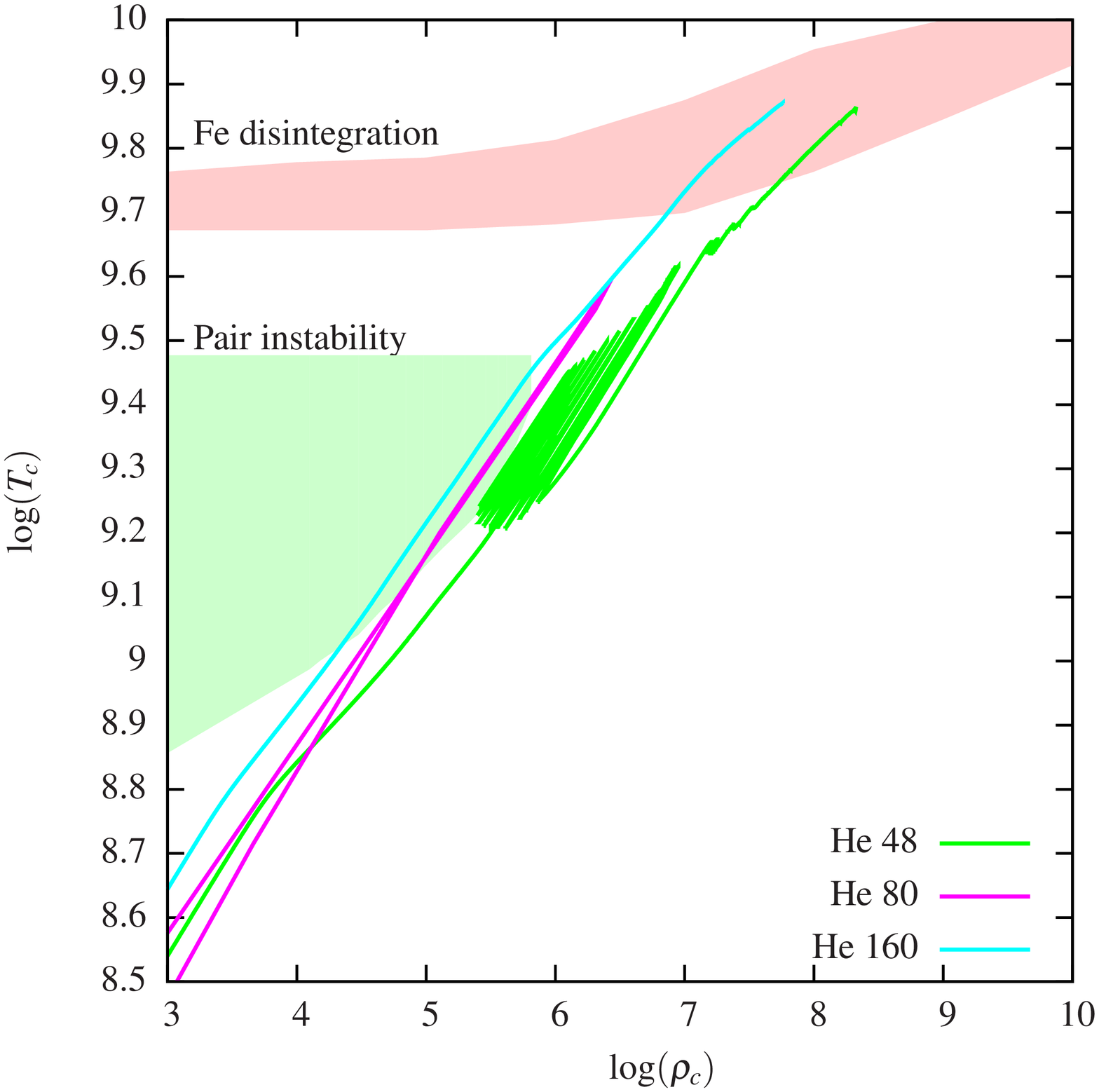}
\end{minipage}
\caption{Evolution of the central density and temperature. Each line
is labeled ``M'' for stellar models and ``He'' for He-core models,
followed by the mass of the model. The figure is divided into two
panels for clarity - the left panel shows models that reach CC
without reaching pair instability, the right panel shows models
reaching pair instability, subsequently experiencing pulsations
(He48), complete disruption (He80), or direct collapse (He160).}
\label{fig:rhoc_tc}
\end{figure}

The He-core models we computed can be divided into four categories,
according to their final fate, as can be seen in the central density
and temperature plot (Fig. \ref{fig:rhoc_tc}):

\begin{enumerate}
  \item CCSN - Models that reach core collapse (i.e. Fe photo-disintegration)
  conditions without entering the region of pair instability. This
  is the fate of He-cores with mass $M \leq 36\,M_\odot$, as can be
  seen for the models He8 and He36 in Fig. \ref{fig:rhoc_tc}.
  \item Pulsational PISN (PPI)- Models that reach pair instability,
  collapse and bounce due to the energy released by nuclear reactions, but the energy released is insufficient
  to disrupt the entire star, thus a fraction of the star's mass is
  emitted, and the star collapses back. This may happen several
  times, until the star has no more material to burn and reverse the collapse, and core
  collapse conditions are reached. This occurs for models with He-core mass in the range
   $36 < M \leq 54\,M_\odot$, e.g. model He48
  in Fig. \ref{fig:rhoc_tc}.
  \item PISN - Models that reach pair instability,
  collapse, and the energy released by nuclear reactions is high
  enough to disrupt the entire star. This occurs for models with He-core mass in the range
   $54 < M \lesssim 130\,M_\odot$, e.g. model He80 in Fig.
  \ref{fig:rhoc_tc}.
  \item Pair instability core collapse (PICC) - Models that reach pair
  instability, but the energy released is too low to reverse the
  collapse, and the star continues collapsing into the
  photodisintegration regime. This occurs for models with He-core mass in the range
   $M \gtrsim 130\,M_\odot$, e.g. model He160 in Fig.
  \ref{fig:rhoc_tc}.
\end{enumerate}

\clearpage
\clearpage

The properties of our models at pre-SN are summarized in table
\ref{tbl:prop}.

\clearpage

\begin{deluxetable}{lrrrrrrrcccccl}
\tabletypesize{\scriptsize}

\rotate

\tablecaption{Properties of the pre-SN models.\label{tbl:prop}}

\tablewidth{0pt}

\tablehead{ \colhead{Model} & \colhead{$M$} & \colhead{$M_{He}$} &
\colhead{$M_{CO}$} & \colhead{$M_{Si}$} & \colhead{$M_{Fe}$} &
\colhead{$X_{C12}$} & \colhead{$X_{Ne20}$}
 & \colhead{$\rho_{c}$} & \colhead{$T_{c}$} & \colhead{$S_{c}$} & \colhead{$Y_e$} & \colhead{$E_{bin}$} & \colhead{SN type}}

\startdata

M20 & 20& 6.1& 2.71& 2.05& 1.44& 0.200& 0.003& 5.9E+09& 7.0& 0.70& 0.424& 0.8& CC\\
M80 & 80& 37.5& 28.20& 10.20& 2.80& 0.050& 0.066& 2.8E+08& 7.0& 1.64& 0.454& 3.9& CC\\
M100& 100& 50.4& 38.90& 20.10& 2.53& 0.037& 0.081& 2.5E+08& 7.0& 1.72& 0.454& 4.0& CC\\
\\
He8 & & 8.0& 5.7& 2.25& 1.62& 0.158& 0.008& 1.1E+09& 7.0& 1.05& 0.440& 1.2& CC\\
He12& & 12.0& 8.8& 3.84& 1.98& 0.124& 0.014& 7.1E+08& 7.0& 1.18& 0.443& 1.7& CC\\
He16& & 16.0& 11.9& 4.01& 1.88& 0.100& 0.021& 5.9E+08& 7.0& 1.25& 0.444& 2.2& CC\\
He20& & 20.0& 15.4& 5.04& 2.01& 0.079& 0.031& 4.9E+08& 7.0& 1.33& 0.447& 2.7& CC\\
He24& & 24.0& 18.6& 6.34& 2.15& 0.072& 0.036& 3.8E+08& 7.0& 1.45& 0.449& 3.1& CC\\
He28& & 28.0& 22.0& 7.99& 2.25& 0.063& 0.044& 3.0E+08& 7.0& 1.60& 0.452& 3.5& CC\\
He32& & 32.0& 25.5& 9.79& 2.18& 0.056& 0.051& 2.3E+08& 7.0& 1.77& 0.456& 3.5& CC\\
He34& & 34.0& 27.3& 11.20& 2.42& 0.053& 0.054& 2.1E+08& 7.0& 1.84& 0.456& 3.6& CC\\
He36& & 36.0& 29.1& 12.10& 2.88& 0.050& 0.057& 2.0E+08& 7.0& 1.91& 0.457& 3.7& CC\\
\\
He38& & 38.0& 30.8& 13.00& 2.67& 0.048& 0.060& 1.8E+08& 7.0& 1.97& 0.457& 3.6& PPI\\
He44& & 44.0& 35.7& 10.80& 2.65& 0.042& 0.069& 2.1E+08& 7.0& 1.86& 0.457& 3.4& PPI\\
He48& & 48.0& 39.1& 17.00& 2.96& 0.039& 0.074& 1.7E+08& 7.0& 2.00& 0.458& 3.1& PPI\\
He50& & 50.0& 40.9& 14.50& 3.05&      &      & 1.8E+08& 7.0& 1.97&
0.459& 4.1& PPI\\
He52& & 52.0& 42.5& 10.50& 2.76&       &      & 1.1E+09& 7.0& 1.03&
0.439& 2.5& PPI\\
He54& & 54.0& 44.3& 18.80& 2.81&       &      & 1.2E+09& 7.0& 1.04&
0.439& 2.9& PPI\\
\\
He56 &  & 56.0 & 46.00 &  &  & 0.034& 0.084& 5.7E+05 & 2.0& & &5.3& PISN\\
He64 &  & 64.0 & 53.00 &  &  & & & 3.7E+05 & 1.9& & &5.8& PISN\\
He72 &  & 72.0 & 60.00 &  &  & & & 2.7E+05 & 1.8 & & &6.4& PISN\\
He80 &  & 80.0 & 67.00 & &  & & & 2.2E+05 & 1.7& & &7.0& PISN\\
He96 &  & 96.0 & 82.00 &  &  & & & 1.5E+05 & 1.6 & & &8.1& PISN\\
He128 & & 128.0 & 110.00 & &  & & & 7.6E+04 & 1.4& & &10.0& PISN\\
\\
He160 & & 160.0 & 144.00 & &  & & & 5.3E+04 & 1.4& & &12.0& PICC\\
\enddata
\tablecomments{The columns represent for each model the total mass
($M$), He-core mass ($M_{He}$), CO-core mass ($M_{CO}$), Si-core
mass ($M_{Si}$), Fe-core mass ($M_{Fe}$), $^{12}C$ and $^{20}Ne$
mass fraction at the end of core He burning ($X_{C12}$ and
$X_{Ne20}$), central density ($\rho_{c}$), temperature ($T_{c}$),
entropy per baryon ($S_{c}$), electron mole fraction ($Y_e$), and
binding energy ($E_{bin}$). Masses are given in $M_\odot$, density
in $g\,cm^{-3}$, temperature in $10^9\,K$, and energy in
$10^{51}\,erg$. For the models reaching core collapse the data are
given when central temperature reaches $7 \times 10^9$, while for
the models that disrupt after reaching pair instability the data are
given at onset of instability.}
\end{deluxetable}

\clearpage

\begin{figure}
\includegraphics[width=1.\linewidth]{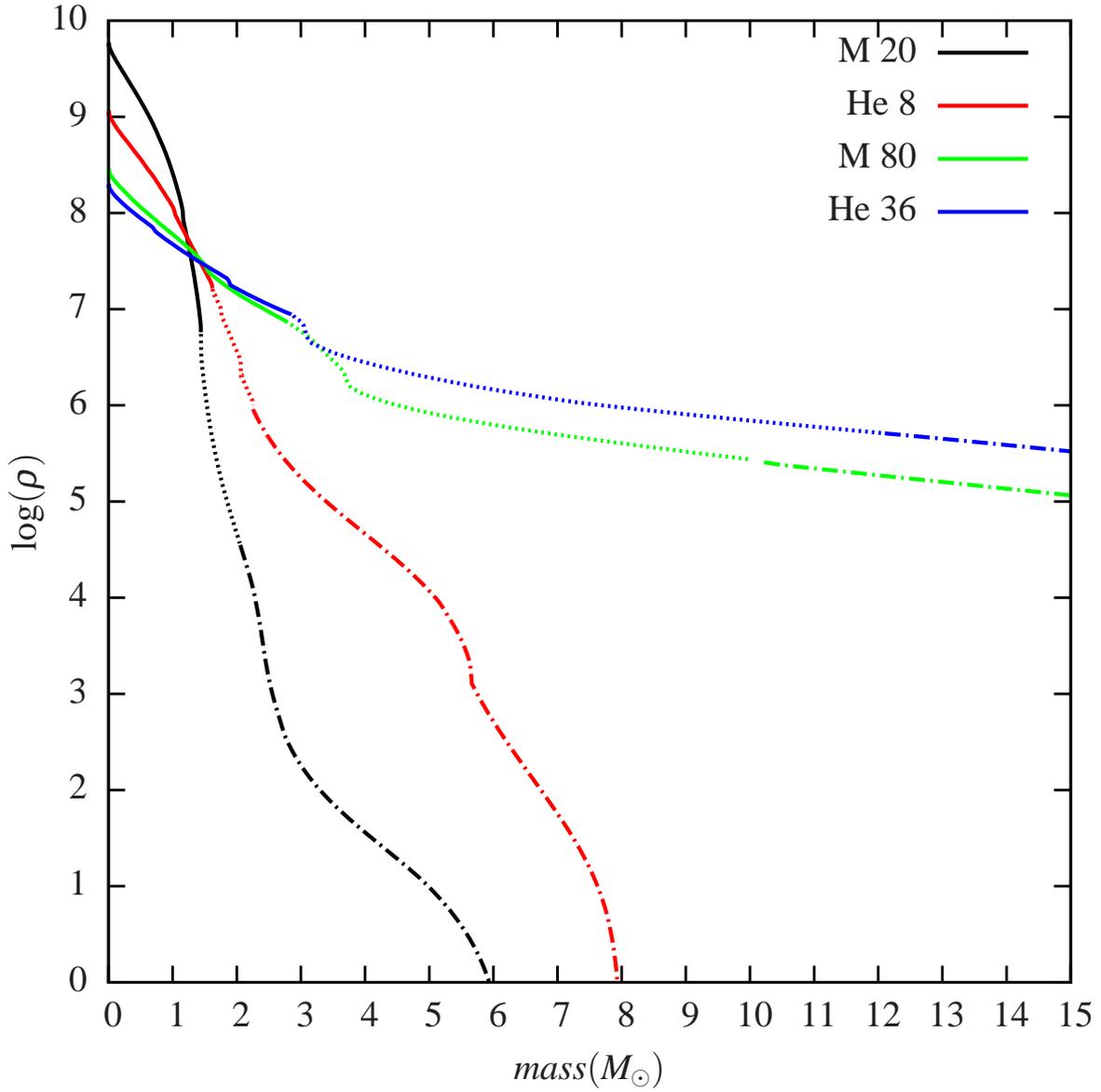}
\caption{Pre-SN density structure. Each line is labeled ``M'' for
stellar models and ``He'' for He-core models, followed by the mass
of the model. (In the color version the solid part of each line
designates the Fe-group core, the dotted part - the Si-group core,
and the dash-dotted - the rest of the model.)} \label{fig:rho_m}
\end{figure}

\begin{figure*}
\begin{minipage}[t]{0.5\linewidth}
\centering
\includegraphics[width=1.\linewidth]{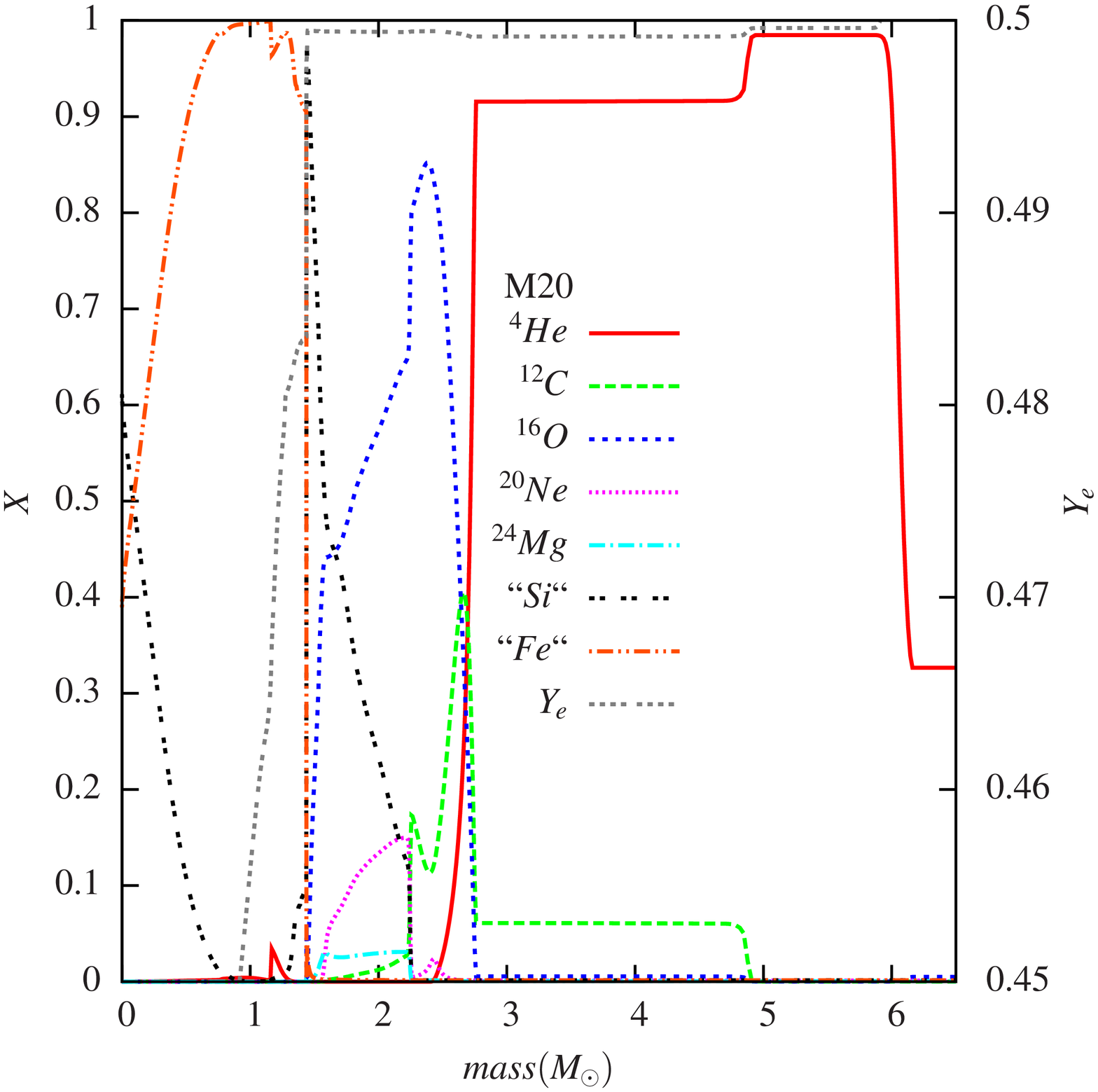}
\end{minipage}
\begin{minipage}[t]{0.5\linewidth}
\centering
\includegraphics[width=1.\linewidth]{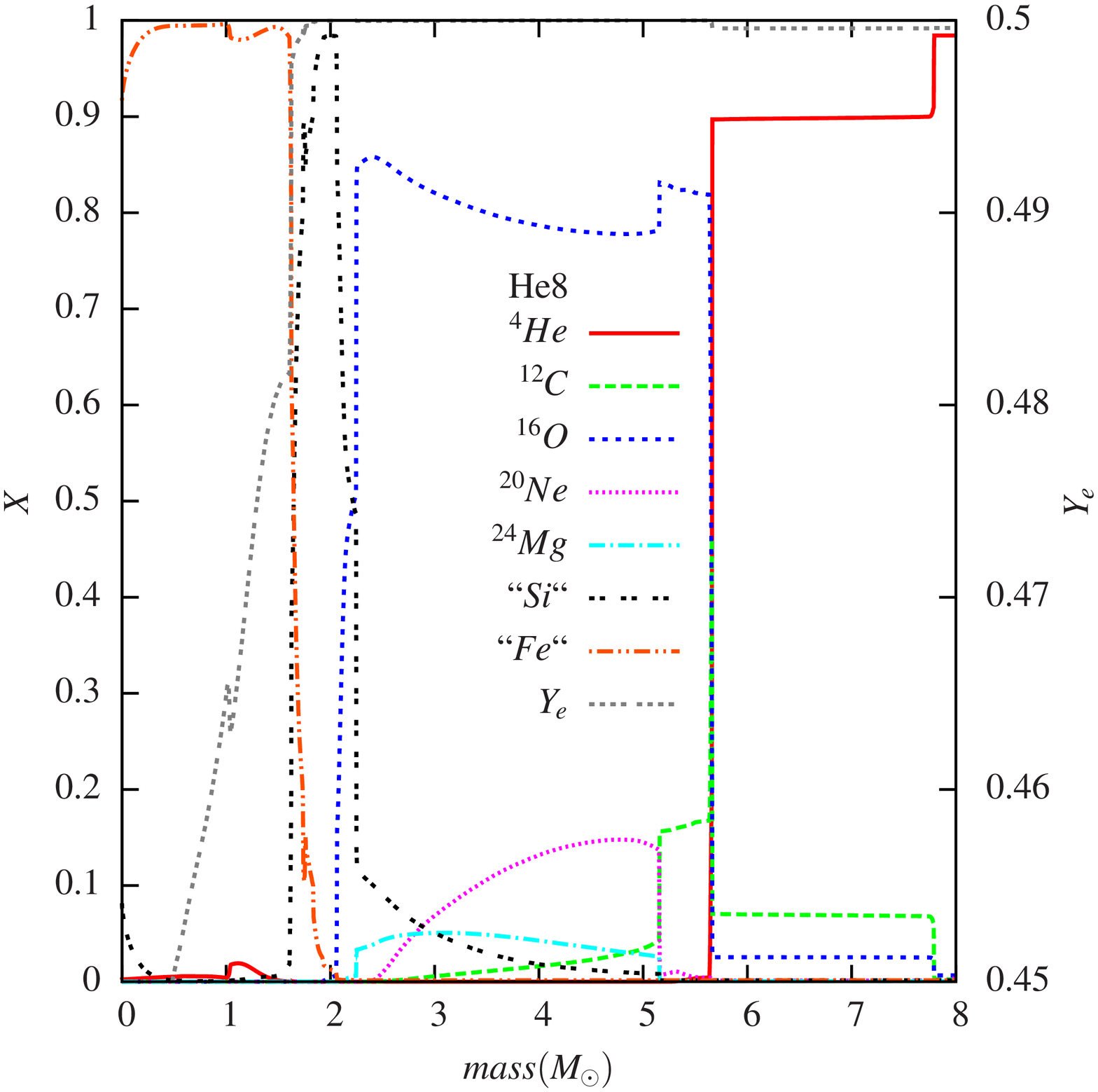}
\end{minipage}
\begin{minipage}[t]{0.5\linewidth}
\centering
\includegraphics[width=1.\linewidth]{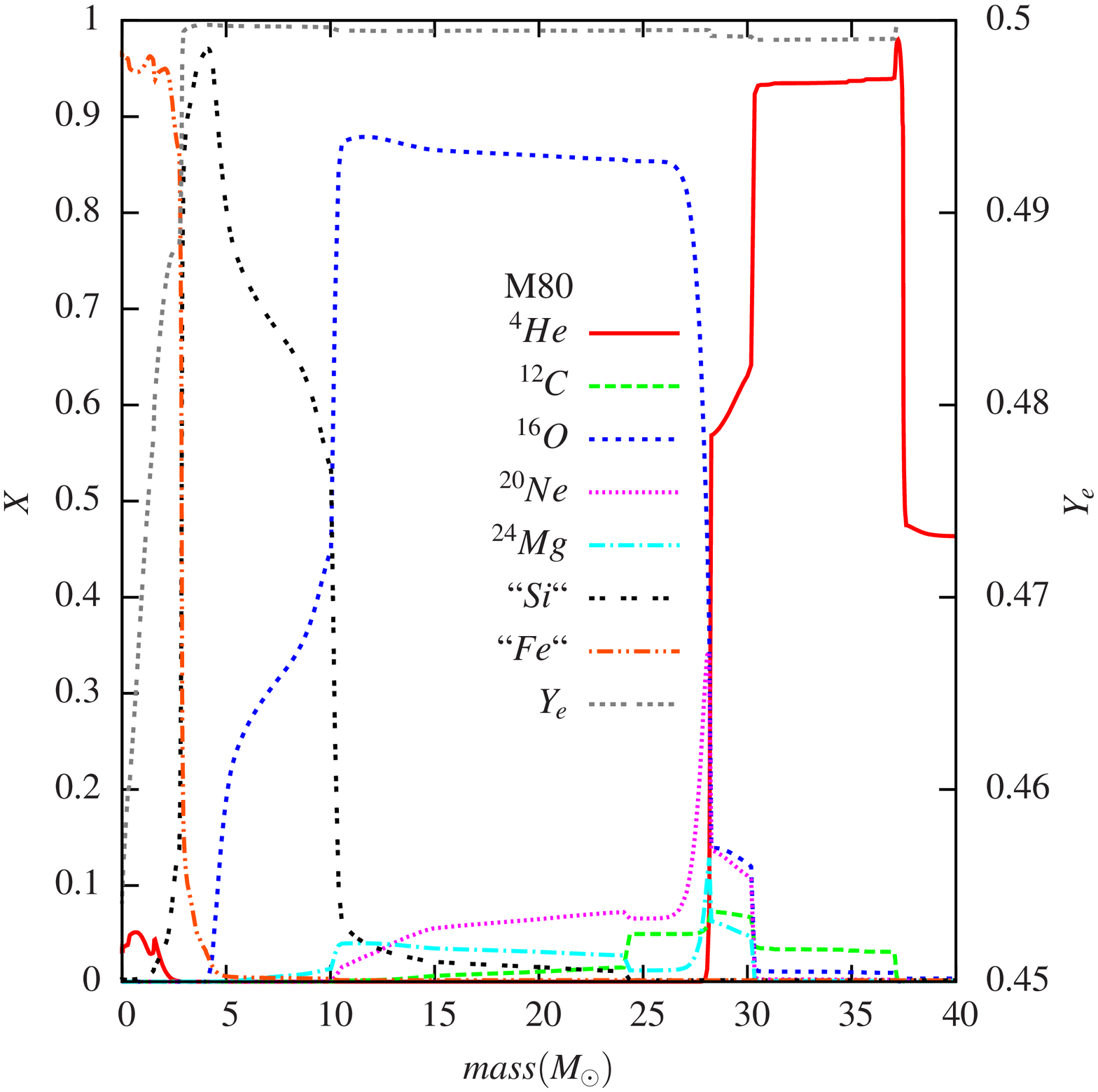}
\end{minipage}
\begin{minipage}[t]{0.5\linewidth}
\centering
\includegraphics[width=1.\linewidth]{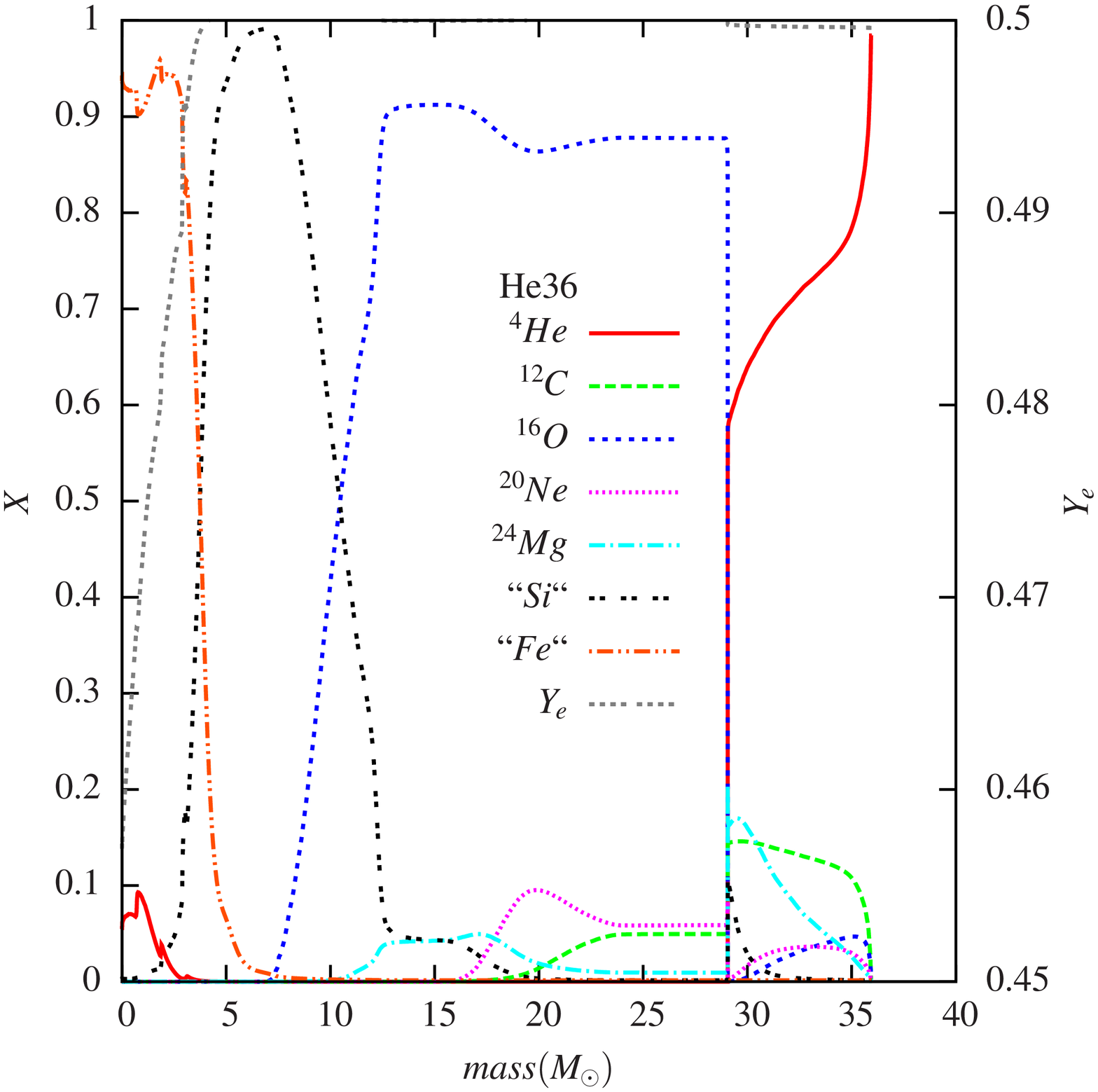}
\end{minipage}
\caption{Pre-SN composition of models M20, He8, M80 and He36. ``Si''
and ``Fe'' stand for the total of Si- and Fe-group elements
respectively.} \label{fig:compos}
\end{figure*}

Fig. \ref{fig:rho_m} shows the density structure of the pre-SN, at
the moment when the central temperature reaches $7 \times
10^{9}\,K$. The two extreme models He8 and He36 are shown, as well
as M80 which has a He-core mass similar to the He36 model, and M20 -
a typical CCSN progenitor. The composition of the same models is
shown in Fig. \ref{fig:compos}.

\clearpage

\clearpage

\begin{figure}
\includegraphics[width=1.\linewidth]{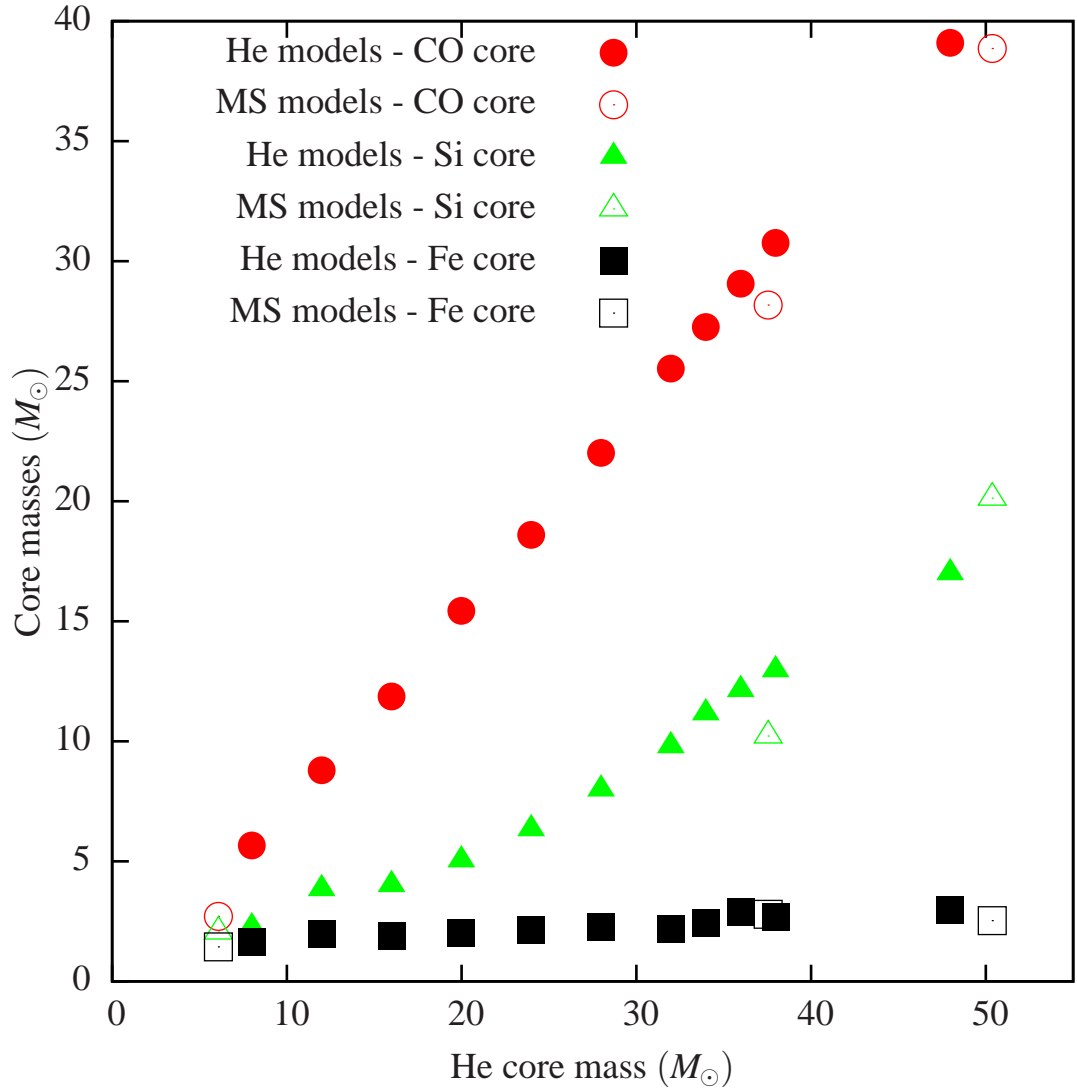}
\caption{Mass of the Fe-core (squares), Si-core (triangles), and
CO-core(circles) for the computed models. Filled shapes designate
He-core models, open shapes - stellar models.} \label{fig:cores}
\end{figure}

\begin{figure}
\includegraphics[width=1.\linewidth]{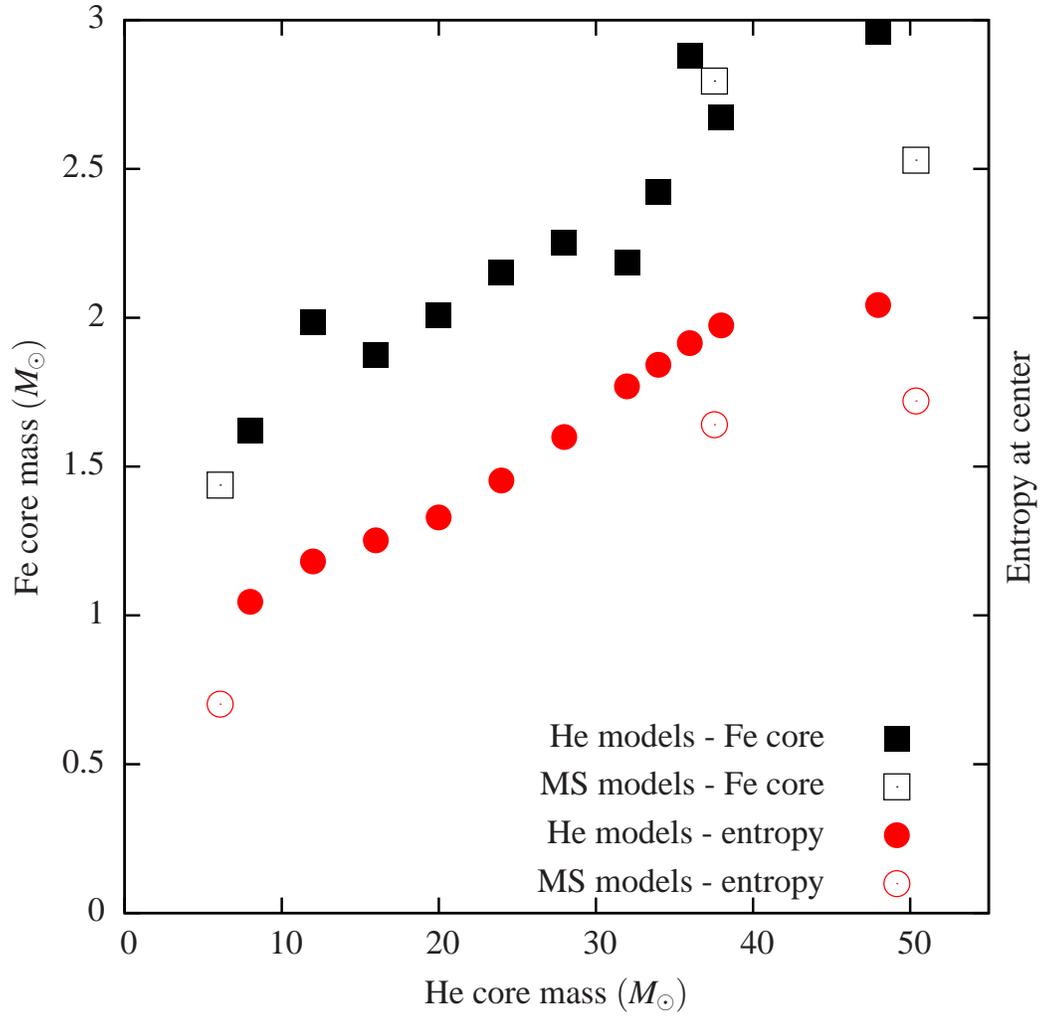}
\caption{Mass of the Fe-core (squares) and central entropy per
baryon (circles) for the computed models. Filled shapes designate
He-core models, open shapes - stellar models.} \label{fig:fe_cores}
\end{figure}

The size of the Fe-, Si-, and CO-cores of our pre-SN models is shown
in Fig. \ref{fig:cores} plotted against the size of the He-core. A
scaled-up view of the size of the Fe-core (defined as the mass
coordinate where the electron mole fraction $Y_e<0.49$) together
with the central entropy per baryon for the same models is shown in
Fig. \ref{fig:fe_cores}. Note that the size of the Fe-core is
slightly non-monotonic. The central entropy, is monotonic with mass,
but slightly differs between He-core and stellar models.

From the above results it is notable that the He-core models behave
similarly to the stellar models (compare e.g. models He36 and M80
which has a He-core mass of $\approx 36\,M_\odot$), however some
differences still exist (e.g. the central entropy in Fig.
\ref{fig:fe_cores}).

\clearpage

\clearpage

\section{Conclusions}

Our results are in general agreement with previously published
results \citep[e.g.][]{Heger2002ApJ...567..532H,
Woosley2007Natur.450..390W, Umeda&Nomoto2008ApJ...673.1014U}. We
focused on the heaviest models which do not encounter pair
instability (CCSN) in the range (He8 - He36). The outstanding novel
features of these models are:
\begin{enumerate}
  \item Relatively large Fe-cores, up to about $3\,M_\odot$, and a
  large amount (up to about $10\,M_\odot$) of Si-group elements.
  \item Comparatively low central density and high central entropy.
  \item A comparatively shallow density profile.
\end{enumerate}

These differences might have a considerable impact on the behavior
of these models during core collapse and on the outcome of the
explosion, a question which we hope to address in the future.

Similar features are encountered for the lower mass part of the
pulsational pair instability models (He38 - He50). However, due to
the numerical complexity of following the pulsations and the
potential sensitivity of the results to the numerical treatment of
convection, mass loss etc., further work is needed to ascertain the
validity of the results for the pulsational models.

\acknowledgements I would like to thank Zalman Barkat for many hours
of fruitful discussion. I would like to acknowledge David Arnett for
providing his TYCHO code for public use.


\begin{thebibliography}{20}
\expandafter\ifx\csname
natexlab\endcsname\relax\def\natexlab#1{#1}\fi

\bibitem[{Barkat {et~al.}(1967)Barkat, Rakavy, \&
  Sack}]{Barkat1967PhysRevLett.18.379}
Barkat, Z., Rakavy, G., \& Sack, N. 1967, Phys. Rev. Lett., 18, 379

\bibitem[{{Bond} {et~al.}(1984){Bond}, {Arnett}, \&
  {Carr}}]{Bond1984ApJ...280..825B}
{Bond}, J.~R., {Arnett}, W.~D., \& {Carr}, B.~J. 1984, \apj, 280,
825

\bibitem[{{El Eid} {et~al.}(1983){El Eid}, {Fricke}, \&
  {Ober}}]{Ober1983A&A...119...54E}
{El Eid}, M.~F., {Fricke}, K.~J., \& {Ober}, W.~W. 1983, \aap, 119,
54

\bibitem[{{Eldridge} \& {Tout}(2004)}]{Eldridge&Tout2004MNRAS.353...87E}
{Eldridge}, J.~J., \& {Tout}, C.~A. 2004, \mnras, 353, 87

\bibitem[{{Fraley}(1968)}]{Fraley1968Ap&SS...2...96F}
{Fraley}, G.~S. 1968, \apss, 2, 96

\bibitem[{{Heger} \& {Woosley}(2002)}]{Heger2002ApJ...567..532H}
{Heger}, A., \& {Woosley}, S.~E. 2002, \apj, 567, 532

\bibitem[{{Heger} \& {Woosley}(2008)}]{Heger&Woosley2008arXiv0803.3161H}
---. 2008, ArXiv e-prints, 803

\bibitem[{{Hirschi} {et~al.}(2004){Hirschi}, {Meynet}, \&
  {Maeder}}]{Hirschi2004A&A...425..649H}
{Hirschi}, R., {Meynet}, G., \& {Maeder}, A. 2004, \aap, 425, 649

\bibitem[{{Langer} {et~al.}(2007){Langer}, {Norman}, {de Koter}, {Vink},
  {Cantiello}, \& {Yoon}}]{Langer2007A&A...475L..19L}
{Langer}, N., {Norman}, C.~A., {de Koter}, A., {Vink}, J.~S.,
{Cantiello}, M.,
  \& {Yoon}, S.-C. 2007, \aap, 475, L19

\bibitem[{{Nakazato} {et~al.}(2006){Nakazato}, {Sumiyoshi}, \&
  {Yamada}}]{Nakazato2006ApJ...645..519N}
{Nakazato}, K., {Sumiyoshi}, K., \& {Yamada}, S. 2006, \apj, 645,
519

\bibitem[{{Nakazato} {et~al.}(2007){Nakazato}, {Sumiyoshi}, \&
  {Yamada}}]{Nakazato2007ApJ...666.1140N}
---. 2007, \apj, 666, 1140

\bibitem[{{Nomoto} {et~al.}(2005){Nomoto}, {Tominaga}, {Umeda}, {Maeda},
  {Ohkubo}, {Deng}, \& {Mazzali}}]{Nomoto2005ASPC..332..374N}
{Nomoto}, K., {Tominaga}, N., {Umeda}, H., {Maeda}, K., {Ohkubo},
T., {Deng},
  J., \& {Mazzali}, P.~A. 2005, in Astronomical Society of the Pacific
  Conference Series, Vol. 332, The Fate of the Most Massive Stars, ed.
  R.~{Humphreys} \& K.~{Stanek}, 374--+

\bibitem[{{Ober} {et~al.}(1983){Ober}, {El Eid}, \&
  {Fricke}}]{Ober1983A&A...119...61O}
{Ober}, W.~W., {El Eid}, M.~F., \& {Fricke}, K.~J. 1983, \aap, 119,
61

\bibitem[{{Ofek} {et~al.}(2007){Ofek}, {Cameron}, {Kasliwal}, {Gal-Yam}, {Rau},
  {Kulkarni}, {Frail}, {Chandra}, {Cenko}, {Soderberg}, \&
  {Immler}}]{Ofek2007ApJ...659L..13O}
{Ofek}, E.~O., {Cameron}, P.~B., {Kasliwal}, M.~M., {Gal-Yam}, A.,
{Rau}, A.,
  {Kulkarni}, S.~R., {Frail}, D.~A., {Chandra}, P., {Cenko}, S.~B.,
  {Soderberg}, A.~M., \& {Immler}, S. 2007, \apjl, 659, L13

\bibitem[{{Rakavy} \& {Shaviv}(1967)}]{Rakavy&Shaviv1967ApJ...148..803R}
{Rakavy}, G., \& {Shaviv}, G. 1967, \apj, 148, 803

\bibitem[{{Rauscher} \&
  {Thielemann}(2000)}]{Rauscher_Thielemann2000ADNDT..75....1R}
{Rauscher}, T., \& {Thielemann}, F.-K. 2000, Atomic Data and Nuclear
Data
  Tables, 75, 1

\bibitem[{{Scannapieco} {et~al.}(2005){Scannapieco}, {Madau}, {Woosley},
  {Heger}, \& {Ferrara}}]{Scannapieco2005ApJ...633.1031S}
{Scannapieco}, E., {Madau}, P., {Woosley}, S., {Heger}, A., \&
{Ferrara}, A.
  2005, \apj, 633, 1031

\bibitem[{{Smith} {et~al.}(2007){Smith}, {Li}, {Foley}, {Wheeler}, {Pooley},
  {Chornock}, {Filippenko}, {Silverman}, {Quimby}, {Bloom}, \&
  {Hansen}}]{Smith2007ApJ...666.1116S}
{Smith}, N., {Li}, W., {Foley}, R.~J., {Wheeler}, J.~C., {Pooley},
D.,
  {Chornock}, R., {Filippenko}, A.~V., {Silverman}, J.~M., {Quimby}, R.,
  {Bloom}, J.~S., \& {Hansen}, C. 2007, \apj, 666, 1116

\bibitem[{{Umeda} \& {Nomoto}(2008)}]{Umeda&Nomoto2008ApJ...673.1014U}
{Umeda}, H., \& {Nomoto}, K. 2008, \apj, 673, 1014

\bibitem[{{Woosley} {et~al.}(2007){Woosley}, {Blinnikov}, \&
  {Heger}}]{Woosley2007Natur.450..390W}
{Woosley}, S.~E., {Blinnikov}, S., \& {Heger}, A. 2007, \nat, 450,
390

\bibitem[{{Young} \& {Arnett}(2005)}]{Young_Arnett2005ApJ}
{Young}, P.~A., \& {Arnett}, D. 2005, \apj, 618, 908

\bibitem[{{Yungelson} {et~al.}(2008){Yungelson}, {van den Heuvel}, {Vink},
  {Portegies Zwart}, \& {de Koter}}]{Yungelson2008A&A...477..223Y}
{Yungelson}, L.~R., {van den Heuvel}, E.~P.~J., {Vink}, J.~S.,
{Portegies
  Zwart}, S.~F., \& {de Koter}, A. 2008, \aap, 477, 223


\end{thebibliography}

\end{document}